\documentclass[preprint,12pt]{elsarticle}

\usepackage{amssymb}

\journal{Physics of the Dark Universe}

\begin{document}
\newcommand{\be}{\begin{equation}}
\newcommand{\ee}{\end{equation}}

\begin{frontmatter}


\title{Turnaround radius in modified gravity}


\author{Valerio Faraoni}
\ead{vfaraoni@ubishops.ca}

\address{Physics Department and STAR Research Cluster, 
Bishop's University, Sherbrooke, Qu\'ebec, Canada J1M 1Z7}

\begin{abstract}
In an accelerating universe in General Relativity there is 
a maximum radius above which a shell of test particles 
cannot collapse, but is dispersed by the cosmic expansion. 
This radius could be used in conjunction with observations 
of large structures to constrain the equation of state of 
the universe. We extend the concept of turnaround radius to 
modified theories of gravity for which the gravitational 
slip is non-vanishing.
\end{abstract}


\end{frontmatter}



\section{Introduction}

Since 1998 cosmologists and theoretical physicists have 
been trying to explain the present acceleration of the 
universe discovered with type Ia supernovae. Apart from the 
problematic cosmological constant, explanations based on a 
dark energy introduced {\em ad hoc} (see 
\cite{AmendolaTsujikawabook} for a review) are not 
satisfactory and many researchers have turned to 
contemplating the possibility of modifying gravity at large 
scales (\cite{CCT}, see \cite{SotiriouFaraoni, 
DeFeliceTsujikawa, NojiriOdintsov, Padilla, 
BakerPsaltisSkordis} for 
reviews). Although modifying gravity is a viable 
possibility, too many dark energy and modified gravity 
models fit the observational data and it is important to 
use any test of gravity which may become available, at all 
scales, to obtain hints on the correct explanation of the 
cosmic acceleration and, possibly, on the correct theory of 
gravity.

One possibility which has been pointed out recently is 
testing theoretical predictions of the turnaround radius 
with astronomical observations \cite{Souriau, Stuchlik1, 
Stuchlik2, Stuchlik3, Stuchlik4, Mizony05, Stuchlik5, 
Nolan2014, DG1, DG2, DG3, DG4, DG5, DG6, Bushaetal, PT, 
PTT, 
TPT}. In an accelerating 
Friedmann-Lem\^aitre-Robertson-Walker (FLRW) universe, 
there is a maximum (areal) radius beyond which a spherical 
shell of dust cannot collapse but expands forever, driven 
by the cosmic accelerated expansion. We formulate our 
final results in terms of the areal radius ($R$) because 
solutions of modified gravity theories are reported in the 
literature using various radial coordinates. 
However, effects peculiar to a particular coordinate system 
would not be meaningful in relativistic 
gravity,\footnote{Unless 
that coordinate system is associated with a preferred 
family of observers.} while  effects 
characterized in a geometric, coordinate-independent, way 
are physically meaningful. The areal 
radius separating two points in space is a physical 
distance 
identified in a completely geometric way by the area of 
2-spheres of symmetry in a spherically symmetric spacetime. 
In a spatially FLRW universe with line element $ds^2=-dt^2 
+a^2(t) \left( dr^2 +r^2 d\Omega_{(2)}^2 \right)$, the 
physical (areal) radius is $R(t,r)= a (t) r$ and it expands 
with the scale factor $a(t)$, while the comoving radius $r$ 
is simply a label attached to elements of the cosmic fluid.

The first comparisons 
of the prediction for the turnaround radius in the 
$\Lambda$CDM model of General Relativity with objects in 
the sky have been carried out \cite{PT, PTT, TPT, 
Bushaetal} and, although the precision is still poor, the 
method holds promise. In General Relativity, the concept of 
turnaround radius can be made more rigorous by using the 
Hawking-Hayward quasi-local mass 
\cite{FaraoniPrainLapierre}. Given the motivation for 
modified gravity in cosmology, here we propose to extend 
the scope of studies of the turnaround radius to 
alternative theories of gravity. We do not commit to any 
specific modified gravity theory at this stage, but adopt a 
post-Friedmannian approach \cite{FerreiraSkordis} in 
which 
a post-FLRW metric fits many theories, to lowest order in 
metric perturbations from an exact FLRW background. 
Contrary to General Relativity, in which two scalar 
potentials in the perturbed FLRW metric coincide (apart 
from the sign), in modified gravity there are two distinct 
potentials which are not trivially related.  We derive the 
turnaround radius in this scheme and find that the physical 
(areal) turnaround radius depends on both potentials. We 
point out that, in order for studies of the turnaround 
radius to be meaningful, it is not sufficient to pick a 
theory of modified gravity but efforts must be made to 
establish which solutions of this theory are generic in 
some appropriate sense.

The metric signature employed in this paper is $-+++$ and 
we use units in which Newton's constant $G$ and the speed 
of light $c$ assume the value unity.

\section{Turnaround radius in the parametrized 
post-Friedmannian approach}

The parametrized post-Friedmannian approach 
\cite{FerreiraSkordis} describes perturbations of a FLRW 
universe in theories of gravity alternative to General 
Relativity. The line element~(\ref{metricCNgauge}) below is 
a rather general parametrization of the metric 
describing perturbed FLRW universes in 
modified gravity 
\cite{Bertschinger2006, JoyceJainKhouryTrodden2015, 
LeonardBakerFerreira2015} (it holds, for example, in 
$f(R)$ gravity \cite{MunshiPrattenValageasColesBrax2015, 
ChiuTaylorShuTu2015}). 
The spacetime metric in the conformal Newtonian gauge is 
\cite{Bertschinger:2008zb, Daniel:2007kk, Daniel:2010ky, 
Pogosian, Zhao:2010dz, Reyes:2010tr}
\be\label{metricCNgauge}
ds^2 = a^2 ( \eta) \left[ -\left( 1+2\psi \right) d\eta^2 +\left( 
1-2\phi \right) \left( dr^2 +r^2 d\Omega_{(2)}^2 \right)\right] \,,
\ee
where $d\Omega_{(2)}^2=d\theta^2 +\sin^2 
\theta \, d\varphi^2$ is the metric on the unit 2-sphere, 
$\eta $ is 
the conformal time (related to the comoving time 
$t$ by 
$dt=ad\eta$), $a(\eta)$ is the scale factor of the spatially flat FLRW 
background, and $\phi$ and $\psi$ are two  post-Friedmannian scalar 
potentials. While in General Relativity it is $\psi =-\phi$, in many 
modified theories of gravity these two potentials do not coincide in 
absolute value and the gravitational slip $\xi \equiv 
\left( \phi-\psi \right) /\phi $ 
is used in experiments aiming at detecting deviations from the standard 
$\Lambda$CDM scenario \cite{Chernin00, 
AmendolaKunzSapone08, Bertschinger:2008zb, 
Daniel:2010ky, Pogosian, Zhao:2010dz, 
Reyes:2010tr}.  The 
ansatz~(\ref{metricCNgauge}) does not include vector and tensor metric 
perturbations, which is justified at lower order for non-relativistic 
velocities, and is common practice in the literature on cosmological 
perturbations\footnote{An effect due to a metric of the form~(\ref{metricCNgauge}) 
would signal modified gravity and, neglecting vector and 
tensor degrees of freedom in the metric (which is 
legitimate at this order of approximation), the potentials 
$\psi$ and $\phi$ are all that remains in the line element. 
However a better characterization of modified gravity than 
the metric is, at least in principle, needed and a rigorous 
derivation of the line element~(\ref{metricCNgauge}) in 
various classes of modified gravity theories is still 
missing. Nevertheless, the turnaround radius will probably 
be unaffected by these theoretical improvements.}  ({\em 
e.g.}, 
\cite{AdamekKunzDurrerDaverio, Adamek2}).

Following the literature on the turnaround radius in cosmology 
\cite{PT, PTT, TPT, Bushaetal}, we assume that the 
FLRW perturbation is spherically symmetric, {\em i.e.}, 
$\phi=\phi(r), \psi=\psi(r)$. The numerical importance of 
deviations from 
spherical symmetry was discussed in \cite{Barrow, PT,PTT}. The 
simplest definition of turnaround radius consists of considering 
spherical shells of test particles respecting the spacetime 
symmetry (that is, expanding or contracting but not shearing nor 
rotating) with areal radius $R$, and imposing that they have 
zero radial acceleration, $\ddot{R} =0$, where an overdot denotes 
differentiation with respect to the comoving time $t$ of the FLRW 
background. We will adopt here this common criterion which, in  
General Relativity, can be justified rigorously  \cite{VMA} 
by making use of 
the Hawking-Hayward quasi-local mass construct  
\cite{Hawking, Hayward}. 
In general, this quasi-local 
energy construct is not defined in modified gravity and here we will 
stick to the simple definition $\ddot{R}=0$ for shells of test 
particles (dust) in radial motion. 

Massive test particles follow timelike geodesics with 4-tangents 
$u^a$ satisfying $u_cu^c=-1$ and the geodesic equation
\be\label{geodesic}
\frac{du^a}{d\tau} +\Gamma^a_{bc}u^b u^c=0 \,,
\ee
which we choose to be affinely 
parametrized by the proper time  $\tau$, and where  
\be\label{Christoffel}
\Gamma^a_{bc}=\frac{1}{2} g^{ad}\left( g_{d 
b,c}+g_{d c,b}-g_{ bc , d} \right)
\ee
are the coefficients of the metric connection. We will perform first 
order calculations in the metric perturbations $\phi$ and $\psi$, but 
the density contrast related to the spatial Laplacian of these metric 
perturbations is allowed to be large \cite{VMA, 
AdamekKunzDurrerDaverio, Adamek2}. 
The normalization
\be
u_cu^c= g_{00}(u^0)^2 +g_{11}(u^1)^2 = -1
\ee
for massive test particles with purely radial motion ($u^2=u^3=0$) 
yields
\be \label{2}
( u^0)^2 = \frac{1}{a^2} \left( 1-2\psi \right) +
\left( 1-2\phi -2\psi \right) (u^1)^2
\ee
to first order. 
Eq.~(\ref{geodesic}) then gives
\be
\frac{du^{\mu}}{d\tau} +\Gamma^{\mu}_{00} (u^0)^2 
+2\Gamma^{\mu}_{01} u^0 u^1 
+\Gamma^{\mu}_{11} (u^1)^2 =0 
\ee
for $\mu=0,1$. Eq.~(\ref{Christoffel}) provides the only  
non-vanishing Christoffel symbols to first order
\begin{eqnarray}
\Gamma^0_{00} &=& \frac{ a_{\eta}}{a} \,,\;\;\;\;
\Gamma^0_{01}=\Gamma^0_{10}=  \psi' \,,\;\;\;\;
\Gamma^0_{11}=  \frac{ a_{\eta}}{a} \left( 1-2\phi-2\psi \right) 
\,,\\
&&\nonumber\\
\Gamma^1_{00} &=& \psi' \,, \;\;\;\;
\Gamma^1_{01}= \Gamma^1_{10}=   \frac{ a_{\eta}}{a}  \,, 
\;\;\;\;
\Gamma^1_{11}=  -\phi '\,,
\end{eqnarray}
where a prime denotes differentiation with respect 
to the radial coordinate $r$ and $a_{\eta} \equiv da/d\eta$. 
Therefore, it is
\begin{eqnarray}
&& \frac{du^0}{d\tau}+  \frac{ a_{\eta}}{a} (u^0)^2 
+2\psi' u^0 u^1 + \frac{ a_{\eta}}{a}  \left( 1-2\phi-2\psi 
\right) (u^1)^2 =0 
\,,\label{8} \\
&&\nonumber\\
&& \frac{du^1}{d\tau}+  \psi' (u^0)^2 + 
\frac{2a_{\eta}}{a} \, u^0 u^1 
-\phi' (u^1)^2 =0 \,. \label{5}
\end{eqnarray}
The areal radius of the spherical spacetime (a geometric  
and coordinate-independent quantity) is
\be
R(t,r)= ar\sqrt{1-2\phi} \simeq ar \left(1-\phi \right)
\ee
to first order. Since
\begin{eqnarray} 
\frac{dR}{dt}&=& \left( \dot{a}r+a\dot{r} \right) \left( 1-\phi  
\right) \,, \\
&&\nonumber\\
\frac{d^2R}{dt^2}&=& \left( \ddot{a}r+2\dot{a}\dot{r} +a\ddot{r} 
\right) \left( 1-\phi  \right) \,, 
\end{eqnarray}
and
\begin{eqnarray}
\dot{r} & \equiv& \frac{dr}{dt}=\frac{dr}{d\eta} \, 
\frac{d\eta}{dt}=\frac{1}{a} \frac{dr}{d\tau}\, 
\frac{d\tau}{d\eta}=
\frac{u^1}{au^0} \,,\\
&&\nonumber\\
\ddot{r} & = &\frac{d}{dt} \left( \frac{u^1}{au^0}\right) = 
-\frac{\dot{a}}{a^2} \, \frac{u^1}{u^0} + \frac{1}{a^2 u^0}\, 
\frac{d}{d\tau} \left( \frac{u^1}{u^0} \right) \,,
\end{eqnarray}
we have 
\be
\frac{d^2R}{dt^2}= \left[ \ddot{a}r+\frac{ \dot{a} u^1}{au^0}+  
\frac{ 1}{ au^0} \, \frac{d}{d\tau} \left( \frac{ u^1}{u^0} \right) 
\right] \left( 1-\phi  \right) \,.
\ee
The criterion $d^2R/dt^2=0$ locating the turnaround radius 
\cite{PT, PTT, TPT} becomes
\be\label{13}
\ddot{a}r+H \, \frac{ u^1}{u^0} +  
\frac{ 1}{ au^0} \left[ \frac{1}{u^0} \,  \frac{du^1}{d\tau} 
-\frac{u^1}{(u^0)^2} \,  \frac{ du^0}{d\tau } \right] =0 \,,
\ee
where $H\equiv \dot{a}/a$ is the Hubble parameter in comoving 
time.  To zero order we have $R=ar$ and 
$\dot{R}=\dot{a}r+a\dot{r}=HR$ (the Hubble law) for timelike 
geodesics, so that 
their 4-tangents have components 
\be\label{c18}
u^{\mu}=u^{\mu}_{(0)}+\delta u^{\mu} =\left(\frac{1}{a}, 0,0,0 
\right)+\delta u^{\mu}
\ee
in coordinates $\left( \eta, r, \theta, \varphi \right)$, where the 
perturbations $\delta u^{\mu}$ are of first order. To 
zero order it is $u^0_{(0)} =d\eta/d\tau=d\eta/dt=1/a$ and 
$ u^1_{(0)} =0$. 
Eq.~(\ref{13}) now yields, to first order, 
\be\label{18}
\ddot{a} r +2\dot{a} \delta u^1 +a \, \frac{d ( \delta 
u^1)}{d\tau}=0 \,.
\ee
Eq.~(\ref{c18}) gives $u^0=a^{-1}$ and $u^1=\delta u^1$ 
which, substituted into eq.~(\ref{5}), yields
\be
\frac{d ( \delta u^1)}{d\tau} +\frac{\psi'}{a^2} 
+\frac{2a_{\eta}}{a^2} \, \delta u^1=0 \,.
\ee
Inserting this equation into eq.~(\ref{18}) and using the 
fact that $a_{\eta}=a\dot{a}$ finally gives
\be
\ddot{a}r -\frac{ \psi'}{a} =0  \label{22}
\ee
to first order. Eq.~(\ref{22}) locates the comoving turnaround 
radius $r_c$ once the post-Friedmannian potential $\psi$ is 
given. In general, this is a trascendental 
equation\footnote{Often in Brans-Dicke and $f(R)$ gravity 
one finds \cite{Brans, CliftonBarrow, Clifton, 
CliftonMotaBarrow} potentials $\psi \propto \left( 
\mu/r\right)^p$ where 
$\mu$ is a constant mass coefficient and the exponent $p$ 
is not integer, which makes eq.~(\ref{22}) non-algebraic.} 
and reduces to an algebraic 
one only for simple forms of $\psi$.
In an accelerating universe ($\ddot{a}>0$), and assuming a 
positive decreasing function $\psi'(r)$ (as is the case, for 
example, for a Newtonian point mass $\psi(r)=-m/r$ and as 
it is reasonable to expect in general for a single spherical 
perturbation), there exists 
a {\em unique} intersection between the straight line of positive 
slope $y=\ddot{a} r$ and the decreasing function $y=\psi'(r)$, 
which defines uniquely the turnaround radius $r_c$. In a 
decelerating universe ($\ddot{a}<0$), by contrast, there is no 
intersection between these two curves and there is no turnaround 
radius.

Eq.~(\ref{22}) is the main result of this paper.  The job is not 
complete, though, because one would like to know the 
{\em areal}, not the comoving, turnaround radius. The areal 
turnaround radius is 
\be 
R_c = a(t) r_c \left[ 1-\phi (r_c)  \right] \,, 
\ee 
where $r_c$ is the unique root of eq.~(\ref{22}). The 
comoving turnaround radius depends only on the 
post-Friedmannian potential $\psi$ but  the areal turnaround 
radius, which is the physical radius, depends also on the second 
potential $\phi$. By expanding this equation as 
\be
r \simeq \frac{R}{a} \left[ 1+\phi \left( \frac{R}{a} \right)\right]
\ee
to first order, eq.~(\ref{22}) turns into the equation satisfied by 
 the areal turnaround radius $R_c$
\be  \label{cc23}
\ddot{a} R_c +\ddot{a} R_c \phi_c -\psi_c'  =0 
\ee
where $\phi_c \equiv \phi \left( R_c/a \right) $ and 
$\psi_c' \equiv \psi' \left( R_c/a \right) $, which makes 
it clear that $R_c$ depends on both 
potentials $\psi$ and $\phi$. To make explicit the 
difference between modified gravity 
and General Relativity one can use the gravitational slip 
$\xi$, which relates the potentials $\phi$ and $\psi$ 
through $\psi=\phi(1-\xi)$. Eq.~(\ref{22}) for the 
turnaround radius $r_c$ in the 
geometry~(\ref{metricCNgauge}) can 
be written as 
\be
\ddot{a}r_c -\frac{ \phi_c' (1-\xi_c)}{a} +\frac{ \phi_c 
\xi_c'}{a}  =0  \,.
\ee
Correspondingly, eq.~(\ref{cc23}) becomes 
\be  
\ddot{a} R_c \left( 1+\phi_c \right) 
-\phi_c' \left( 1-\xi_c \right) 
+\phi_c \xi_c ' =0 \,.
\ee

\section{Examples}

Here we consider a few examples. Unfortunately, only a 
handful of analytical solutions of modified gravity 
theories are known which describe spherically symmetric 
inhomogeneities embedded in FLRW universes, and for these 
spacetimes the cosmological ``background'' is invariably 
decelerated instead of accelerated (see the recent review 
\cite{mylastbook}), so they are not useful here.

\subsection{$\Lambda$CDM model}

As a first example, consider the $\Lambda$CDM model of 
General 
Relativity and the simple spherical perturbation potentials 
$\psi(r)=-\phi(r)= -m/r$, with 
$m$ a mass constant. This expression makes sense for 
sufficiently large values of $r$ in order to keep the 
metric perturbations small.  Using the FLRW acceleration 
equation
\be
\frac{ \ddot{a}}{a}= -\frac{4\pi }{3} \left( 
3P+\rho\right) \,,
\ee
where $P$ and $\rho$ are the pressure and energy density 
of the dark effective fluid dominating the dynamics of the 
accelerated universe, respectively,  eq.~(\ref{22}) 
provides the comoving turnaround radius
\be
r_c=\left[ \frac{ 3m}{4\pi \left| \rho +3P \right| a^2} 
\right]^{1/3} 
\,.
\ee
The corresponding areal radius is, to dominant order, 
\be
R_c=ar_c= \left( \frac{ 3ma }{4\pi \left| \rho +3P \right|} 
\right)^{1/3} 
\ee
where the comoving scale $ma$, rather than the constant 
$m$, is the physical mass to consider (see the 
corresponding mass in 
Refs.~\cite{PT, PTT, TPT} and the discussion in \cite{VMA}). 
This equation reproduces 
previous expressions of the turnaround radius found in 
Refs.~\cite{Souriau, Stuchlik1, 
Stuchlik2, Stuchlik3, Stuchlik4, Mizony05, Stuchlik5, 
Nolan2014, DG1, DG2, DG3, DG4, DG5, DG6, Bushaetal, PT, 
PTT, 
TPT}. In an expanding 
universe, the quantity $ma$ is identified with the total mass 
enclosed in a sphere of comoving radius $r$ and areal radius $R=ar$. 
This mass comprises the contribution of the ``local'' perturbation plus 
that of the cosmic fluid enclosed in the sphere  \cite{PT, 
PTT, VMA}. 
In General Relativity, this mass $ma$ 
coincides with the Hawking-Hayward quasi-local energy 
\cite{Hawking, Hayward} and, because 
of spherical symmetry, also with the Misner-Sharp-Hernandez 
mass \cite{Hayward}. In modified gravity, however, these 
quasi-local energy constructs are not defined.

\subsection{Generalized McVittie attractor}

Generalized McVittie solutions of the Einstein equations 
were introduced in \cite{FaraoniJacques}. The old McVittie 
class of solutions of General Relativity \cite{McVittie} 
describes a central object (under certain conditions, a 
black hole) embedded in a FLRW universe. The McVittie class 
of spacetimes has been the subject of much recent attention 
\cite{Kleban, Lake1, Lake2, DSFG2012, Roshina1, Roshina2, 
AndresRoshina, DaSilvaGuarientoMolina2015} and it has been 
shown to solve the equations of cuscuton theory, a special 
Ho\v{r}ava-Lifschitz theory \cite{NiayeshDaniel}. Accretion 
of 
the surrounding cosmic fluid onto the central McVittie 
object is artificially forbidden: in \cite{McVittie}, 
McVittie imposed that the time-radius component $G_{tr}$ of 
the Einstein tensor vanishes, corresponding to zero radial 
flux of mass-energy onto the central inhomogeneity. The 
rationale for this assumption was simplifying the search 
for a solution of the Einstein equations describing a 
central object (in McVittie's intentions, a point mass) 
embedded in a cosmological space. McVittie realized that it 
was much simpler to forbid accretion than having to 
model it. Nonetheless, in 
General Relativity, one can allow a radial flow of energy, 
thus obtaining the so-called ``generalized McVittie 
solutions'' \cite{FaraoniJacques}. This radial 
energy flow 
is spacelike and rather artificial in this context. All 
simple models of heat conduction or of viscosity of 
imperfect fluids in relativity suffer from the  
inconsistency that heat conduction is instantaneous and 
obviously contradicts relativity, but imperfect fluids are 
used routinely as simple models because a proper 
relativistic thermodynamical treatment increases the 
complication to the point of rendering simple models 
impossible. Within the context of 
Horndeski gravity, the generalized McVittie solutions 
become more realistic scalar field solutions 
\cite{Horndeski}, in which the scalar field is suceptible 
of an effective fluid description. Within the class of 
generalized 
McVittie solutions there is a late-time attractor 
\cite{FaraoniGaoetc}, with geometry described by
\be\label{McVattractor}
ds^2 =-\left( 1-\frac{2m_0}{r} \right) dt^2 +a^2(t) \left[ 
\frac{dr^2}{1-\frac{2m_0}{r}} +r^2 d\Omega_{(2)}^2 \right] \,,
\ee
where $m_0$ is a constant. In the approximation $m_0/r \ll 1$, this  
line element reduces to 
\be
ds^2 =-\left( 1-\frac{2m_0}{r} \right) dt^2 +a^2(t) \left[ 
\left( 1+\frac{2m_0}{r}  \right)dr^2 +r^2 d\Omega_{(2)}^2 \right] 
\ee
with $\psi=-m/r=-\phi$ and we reduce formally to the previous 
example. This relation is not surprising since the 
metric (\ref{McVattractor}) solves both the Einstein  and 
the Horndeski field equations and the ansatz 
(\ref{metricCNgauge}) represents a {\em solution} of the field 
equations of  a theory, not the theory itself. Therefore, in this case, 
the turnaround radius does not discriminate between the 
$\Lambda$CDM model based on General Relativity and Horndeski 
gravity.

\section{Discussion and conclusions}

In generalized gravity, when the line element assumes the 
post-Friedman\-nian form~(\ref{metricCNgauge}) widely used 
in 
the literature and with a presently accelerating FLRW 
background, we have found a formula (eq.~(\ref{22})) for 
the turnaround radius $R_c$. The upper bound to the radius 
of stable virialized structures that can be obtained by 
observing the largest bound objects in the universe (those
that are on the verge of breaking apart) would constitute 
an observable to look at in order to see a signature. 
Indeed the potentials $\phi$ and $\psi$ are 
commonly regarded as observables in modified gravity 
\cite{MottaSawickySaltasAmendolaKunz2013, 
BelliniSawicky2014}.

Several modified gravity theories admit multiple 
representations, the most well known situation being that 
of scalar-tensor gravity which can be analyzed in two 
conformal frames, the so-called Jordan frame and Einstein 
frame related by a conformal transformation of the metric 
plus a 
non-linear redefinition of the Brans-Dicke-like scalar 
field present in the theory \cite{Dicke, FujiiMaedabook, 
myfirstbook}. These two frames are in principle physically 
equivalent provided that a rescaling of the units of mass, 
length, and time (and of all the derived units) is 
performed (\cite{Dicke}, see also 
\cite{Flanagan, FaraoniNadeau} 
and the references therein). However, it may be rather 
cumbersome to relate physical quantities in the two frames 
(see, {\em e.g.}, the recent discussion 
\cite{FaraoniPrainZambrano}), which renders the equivalence 
less useful for practical purposes. 

Before eq.~(\ref{22}) can be 
compared with data on observed structures which are 
presumably on the verge of virialization (as done in 
Refs.~\cite{PT, PTT, TPT}), some more work needs to be 
done. Eq.~(\ref{22}) may test how accurately
 eq.~(\ref{metricCNgauge}) describes the spacetime geometry 
to first order in the metric perturbations, and it is 
unlikely that observations will require higher precision 
than first order. However, the same spacetime metric may be 
a solution of several theories of gravity, including 
General Relativity, at once. Thus, the theoretical problem 
arises of which solutions of a given theory of gravity are 
significant, in the sense of being generic and of being 
different from the standard one of Einstein's theory. There 
is some degeneracy, in the sense that different theories of 
gravity, including General Relativity, can admit the same 
post-Friedmannian metric as a solution to first order in 
the metric perturbations (the same way that most theories 
of gravity admit the FLRW or the Schwarzschild metric as an 
exact solution). If, in addition to the FLRW background, 
the perturbed solution is the same then observational 
constraints on the turnaround radius will not detect 
deviations from General Relativity even if gravity was 
described by an alternative theory. Work needs to be done 
to understand which perturbed cosmological solutions are in 
some sense ``generic'' and, therefore. expected in various 
theories of gravity and whether the turnaround radius 
computed for such theories differs significantly from that 
of Einstein's theory. If modified gravity does indeed 
explain the cosmic acceleration without dark energy, as is 
hypotesized, then it deviates from Einstein gravity at 
large scales and the turnaround radius holds promise for an 
independent test of gravity at the interface between 
astrophysics and cosmology.

A completely different, statistical mechanics approach to 
the turnaround radius is contained in 
Refs.~\cite{AxenidesGeorgiuRupas2012, 
RoupasAxenidesGeorgiusaridakis2012}, in which galaxy 
clusters 
are studied by including effective ``dark energy 
particles'' forming a perfect fluid and it is found that 
there is a ``reentrant 
radius''  corresponding to our turnaround radius. Beyond 
this radius thermal equilibrium of the gas is not possible 
and quintessence (with $P>-\rho$) increases the 
gravitational instability, while phantom energy (with 
$P<-\rho$) helps stabilizing against gravitational 
collapse. A comparison betwen our macroscopic approach and 
the statistical mechanics approach of 
\cite{AxenidesGeorgiuRupas2012,
RoupasAxenidesGeorgiusaridakis2012} is not straightforward 
because these authors assume a barotropic equation of 
state of state for dark energy which is linear and 
constant. While it is possible to write the field equations 
of modified gravity in the form of effective Einstein 
equations by collecting geometric terms into the right hand 
side as an effective fluid, in general the latter is not a 
perfect fluid and the effective equation of state is not 
linear nor constant. Imposing a constant effective equation 
of state would results in unphysical restrictions on the 
theory of gravity motivated only by mathematics. 
Nevertheless, it will be interesting to generalize the 
approach of \cite{AxenidesGeorgiuRupas2012,
RoupasAxenidesGeorgiusaridakis2012} to modified gravity in 
the future.



\section*{Acknowledgments}
We thank two referees for helpful comments. This work is 
supported by the Natural Sciences and Engineering Research 
Council of Canada and by Bishop's University.



\end{document}